\documentclass[aps,prd,twocolumn,superscriptaddress]{revtex4}
\usepackage{graphicx}
\usepackage{amsmath,amssymb,amsfonts}
\usepackage[colorlinks,citecolor=blue,linkcolor=blue,urlcolor=blue]{hyperref}

\usepackage[english]{babel}

\usepackage[autostyle, english = british]{csquotes}
\MakeOuterQuote{"}

\begin{document}

\title{The notion of locality in relational quantum mechanics}

\author{P. Martin-Dussaud}
\email{pmd@cpt.univ-mrs.fr}
\affiliation{Aix Marseille Univ, Universit\'e de Toulon, CNRS, CPT, Marseille, France}
\affiliation{Basic Research Community for Physics}
\author{C. Rovelli}
\affiliation{Aix Marseille Univ, Universit\'e de Toulon, CNRS, CPT, Marseille, France}
\author{F. Zalamea}
\affiliation{Basic Research Community for Physics}
\affiliation{EFREI Paris}

\date{\small\today}

\begin{abstract}
\noindent The term "locality" is used in different contexts with different meanings. There have been claims that relational quantum mechanics is local, but it is not clear then how it accounts for the effects that go under the usual name of quantum non-locality. The present article shows that the failure of "locality" in the sense of Bell, once interpreted in the relational framework, reduces to the existence of a common cause in an indeterministic context. In particular, there is no need to appeal to a mysterious space-like influence to understand it.
\end{abstract}

\maketitle 

\section{Notions of locality} \label{1}

It has often been argued that non-locality is a fundamental feature of the world that quantum mechanics has unveiled \cite{Bell1975, Stapp1975}, but it is not clear what this means precisely. The term "locality" is used in different contexts with different meanings. One encounters at least five different notions of "locality" in the literature:
\begin{enumerate}
\item No superluminal signalling: signals cannot propagate faster than light;
\item No superluminal causal influence: causes and effects of events are no further away than permitted by the velocity of light;
\item No space-like influence: space-like separated quantum systems do not influence each other;
\item Point-like interaction: quantum systems (or fields of the Lagrangian in quantum field theory) interact only at the same point in spacetime;
\item Local commutativity: space-like separated local observables commute.
\end{enumerate}
Of course, these various notions can be very close to one another. Sometimes, they are even seen as equivalent, but in the following we will see them as distinct notions, whose meaning will be clarified in due course.

In the 1970s John Bell proposed a precise mathematical formalisation of the principle that we called above "no superluminal causal influence". This formal definition, which originally goes under the name of "local causality", is a crucial step toward the proof of his famous theorem \cite{Bell1990}. When people say that EPR-type experiments highlight the fundamental non-locality of nature, they implicitly refer to that peculiar notion of locality.

In the context of the relational interpretation of quantum mechanics \cite{Rovelli1996}, there have been claims that quantum mechanics becomes local \cite{Smerlak2007}. It is not clear, however, how this presumed "locality" has to deal with the original definition of Bell. Surely, the relational interpretation must have a way to include the effects that go under the usual name of quantum non-locality. This has been recently pointed out as one of the open problems of the relational interpretation \citep{Laudisa2017b}.

The present article proposes a reinterpretation of the mathematical definition of locality in the light of relational quantum mechanics, and clarifies in what sense relational quantum mechanics can be said to be local.

In section \ref{2} we review the notion of "local causality" and its mathematical definition introduced by Bell. In section \ref{3} we see that this definition still makes sense in the relational interpretation, but its meaning changes to the point that there is nothing more surprising in non-locality than in the fundamental randomness of quantum mechanics.

\section{Is quantum mechanics locally causal?} \label{2}

{\bf The failure of local determinism}

For John Bell, a theory is said to be \textit{locally deterministic} if the state of physical systems in a bounded region of space-time $A$ can be entirely deduced from the knowledge of the state of systems in another bounded region $B$ located inside the past light-cone of $A$ \cite{Bell1975}. One can say for instance that Maxwell's theory of electromagnetism is locally deterministic, because one can predict the configuration of electromagnetic fields in a bounded regions of spacetime $A$ knowing the configuration of the fields over a time-slice of the past light-cone of $A$. A counter-example is quantum mechanics, which is not locally deterministic.

It is important to notice that all the indeterminism of quantum mechanics lies in the measurement process. If one restricts to the unitary evolution, given by the Schr\"odinger equation, then the evolution is deterministic (and even more deterministic than in classical mechanics as pointed out by John Earman \cite{Earman2007}). But once a measurement takes place, the future outcome is not determined by past measurements. 

Nevertheless, even if the past does not completely determine the future, there might still be some sense in which one could say that "the future is only influenced by past (and not by spatially separated events)". This is another way to state what we previously called the principle of "no superluminal causal influence". In this way, the notion of locality first appears as an attempt to preserve the intuitive notions of cause and effect in an indeterministic context. John Bell proposed a formalisation of this idea that he called "local causality". Importantly for what follows, this definition relies on the notion of "local beables".  As we shall argue below, quantum non-locality is strictly connected to what we count as local beable.

\vspace{1 \baselineskip}

{\bf Local beables}

Although the notion of beables seems to refer to a peculiar interpretation of quantum mechanics, or even a new theory (one of the first papers by Bell was entitled, maybe misleadingly, \textit{The theory of local beables} \citep{Bell1975}), the concept is meaningful in ordinary quantum mechanics. To put it in a nutshell, the term "beable" is a fancy way to say "element of reality". John Bell has been more or less explicit about what he really meant:
\begin{quote}
\textit{The beables of the theory are those elements which might correspond to elements of reality, to things which exist. Their existence does not depend on "observation". } \cite{Bell1984}
\end{quote}
It was a surprise of quantum mechanics to realise that some very intuitive concepts like "the position of a particle" may simply not be meaningful in the absence of an observation. Reality seemed to fade suddenly. The introduction of the concept of "beable" was an attempt to bring back to the theory the primacy of "things which really \textit{are} in the world" over "things which are observed" (observable). Thus, the concept of beable strongly depends on a certain form of realism, which might first seem to be too restrictive for a good interpretation of quantum mechanics. Is it indeed very reasonable to assume the existence of "elements of reality" in the quantum world? 

In fact, this apparent drawback can be turned into an advantage if beables are seen as a very general concept whose actual content depends on the choice of the interpretation. The ontology of quantum mechanics is not given a priori by its mathematical formalism, but it is expected from the interpretation to supplement the mathematics and to answer the question "what is real?" or equivalently "what does physically exist?". The answer to these questions constitutes the actual content of the term "beable", which is thus a word whose precise meaning may change with the various possible ontologies of quantum mechanics.

Different interpretations may agree on some of the basic things that should be considered as "real". For instance, John Bell suggested that:
\begin{quote}
\textit{The beables must include the settings of switches and knobs on experimental equipment, the current in coils, and the reading of instruments. "Observables" are} made, \textit{somehow, out of beables.} \cite{Bell1975}
\end{quote}

In the following, we will focus on local beables, that is to say on beables that are localized in a bounded region of spacetime. For this to be meaningful, spacetime is assumed to be the classical spacetime of special relativity, upon which "elements of reality" can live. This is not surprising since it is what is usually assumed in ordinary quantum mechanics or in quantum field theory, but one should note that it is still unclear what "local beable" would mean in a context of quantum gravity where spacetime itself would be considered as a quantum field.

The notion of local beables, which we have just seen to be quite flexible, is the basic concept for Bell's formalisation of the principle of "no superluminal causal influence" according to which causes and effects cannot propagate faster than light.

\vspace{1 \baselineskip}

{\bf Local causality}

\begin{figure}[h!]
\includegraphics[width = 1 \columnwidth]{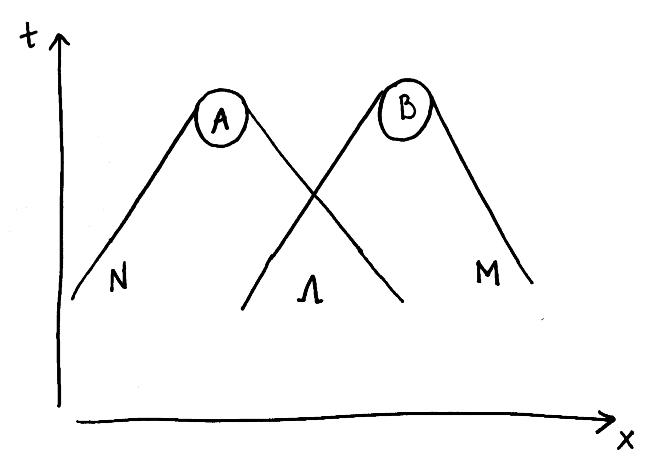} 
\caption{Spacetime diagram showing the localisation of the beables $A$,$B$, $\Lambda$, $N$ and $M$.}\label{diagram1}
\end{figure}

Let us denote $ \{ x | y \}$ the probability of some particular value $x$ of the beable $X$, knowing the particular value $y$ of the beable $Y$. Let $A$ and $B$ be two space-like separated beables, $\Lambda$ be the set of beables in their common past, and $N$ (resp. $M$) the set of beables in the past of $A$ (resp. $B$) excluding $\Lambda$ (see Figure \ref{diagram1}).

Then the theory is said to be \textit{locally causal} if, for all possible values $a, \lambda, n, b$ of the beables $A, \Lambda, N$ and $B$, we have
\begin{equation}
\left\{ a | \lambda, n, b \right\} = \left\{ a | \lambda, n \right\}.
\end{equation}
In words, it means that knowing the value of the beable $B$ does not add any new information that would not be already contained in the beables of the past cone of $A$.

\vspace{1 \baselineskip}

{\bf Quantum mechanics is not locally causal.}

Bell emphasized that quantum mechanics is not locally causal. To see it, consider a radioactive particle in the region $\Lambda$. The radioactive decay will lead to the emission of an $\alpha$ particle. Suppose the beables $A$ and $B$ are the reading of Geiger counters which tell us whether an $\alpha$ particle is detected (value $1$) or not (value $0$). Suppose also that the radioactive particle can only emit at most one $\alpha$ particle. Knowing only the values $\Lambda = \lambda$ and $N = n$, it is impossible to predict if $A$ will detect an $\alpha$ particle or not: it is the randomness of the measurement process. So whatever the value $a \in \{0,1\}$ of the beable $A$, we have
\begin{align*}
\left\{ a | \lambda, n \right\} = \frac 12.
\end{align*}
Now, if $B$ detects an $\alpha$ particle, $A$ will not, and so:
\begin{align*}
\left\{ A=1 | \lambda, n , B=1 \right\} = 0.
\end{align*}
Thus, in this scenario, we have exhibited a case where: 
\begin{align*}
\left\{ a | \lambda, n, b \right\} \neq \left\{ a | \lambda, n \right\}.
\end{align*}
Hence quantum mechanics is not locally causal.

One sees that it is not necessary to introduce EPR pairs to show that ordinary quantum mechanics (assumed complete) is not locally causal. In fact, Einstein, Podolsky and Rosen developed their arguments the other way around. They started by assuming locality and then arrived to the conclusion that quantum mechanics was incomplete \cite{Einstein1935}. Consequently, the hope was to embed quantum mechanics into a larger theory which would be complete and locally causal (the so-called local hidden variables theories). EPR pairs have later been used in experiments to discard such a possibility \cite{Aspect1982}. Quantum mechanics is thus acknowledged to be non-local in the sense of Bell, whether or not it is complete. Therefore, discussing the meaning of non-locality does not require a priori to introduce EPR pairs. It is sufficient to consider the simpler example of radioactive decay as we did.

\vspace{1 \baselineskip}

{\bf Interpreting non-locality}

The usual interpretation of non-locality is fuzzy. If "local causality" is indeed a faithful mathematical instantiation of the principle of "no superluminal causal influence", then its experimental violation should be understood as the possibility that causes and effects propagate faster than light. Furthermore, one usually stresses that this process cannot be used to transmit information, hence no violation of the principle of "no superluminal signalling", which would have been in complete contradiction with special relativity.

However, even the simple idea of superluminal causal influence is at odds with special relativity. One often says that the outcome of a measurement in A determines a later outcome in B, but this can only be a loose way of speaking because the same ensemble of space-like separated outcomes in A and B can be equivalently interpreted as a measure in A affecting B, or a measure in B affecting A, depending on the choice of reference frame specifying a preferred time foliation. The absence of an absolute time ordering between A and B prevents us from interpreting the origin of the correlations by a causal influence from A to B or from B to A, because "causation" is a time-oriented concept. One could argue in favour of an absolute reference frame which would justify an absolute causal orientation, but this hypothesis does not show up in the phenomenology of the experiments. So, the hypothetical non-local influence between A and B cannot be causally oriented. At best, it can be thought as a kind of mutual action at a distance that would enable "instantaneous" space-like influence, and so would violate what was called earlier the principle of "no space-like influence". 

Some other interpretations of non-locality argue that the mere collapse of the wave function would already be a manifestation of non-locality in ordinary quantum mechanics. This is not completely true, because the definition of Bell is a bit more subtle. In fact, it is true if one adopts an ontic interpretation of the wave-function, that is if one regards it as a beable. It is the case, for instance, in Bohmian mechanics. But if one sees the wave-function only as a mathematical trick, then the so-called collapse of the wave function does not tell us anything deep about the local causality of the world. This kind of fake non-locality is very similar to the apparent superluminal propagation of the potential for Maxwell's theory in Coulomb gauge. The same thing also happens for British sovereignty since the Prince becomes the King as soon as the Queen dies, however far away in the Universe he may be. These two examples were exhibited by Bell himself. In both cases, the thing (the potential or the sovereignty) that travels faster than light is not a physical thing. This explains the necessary use of "beables" to define a physically meaningful notion of "locality". By the way, this remark also discards some interpretations of non-locality claiming that EPR-type experiments would force us to choose between "locality" and "realism", for there is no physically meaningful notion of "local causality" without the realism of beables.

The definition of local causality is mathematically well-defined, however its meaning is not obvious because it supervenes on that of beables, which makes it dependent on the interpretation of quantum mechanics. In the following section we are going to see that relational quantum mechanics can still make sense of this definition, but it loses on the way most of its surprising features.

\section{Relational local causality} \label{3}

The relational interpretation of quantum mechanics was proposed in 1995 \cite{Rovelli1996, Rovelli2018}. First of all, it is a criticism of the usual notion of "quantum state". The "state" should not be taken as an absolute notion, i.e. observer independent, but rather as a book-keeping device relative to a specific observer. In this interpretation, the apparent paradox of Wigner's friend is taken as evidence that different observers may give different accounts of the same sequence of events. 

This point of view might be puzzling at first sight because it challenges the opinion that physical systems should be describable independently of any observer. This is surely a departure from strong realism where, for instance, the property "spin up" belongs to the electron independently of any observer. Actually, the Kochen-Specker theorem has already challenged the strongly realistic point of view, showing that a complete set of properties cannot be consistently attributed to a physical system \cite{Kochen1968}. The relational interpretation challenges strong realism a bit differently though: there would be no absolute state-property of a physical system at all. However, this view is not completely anti-realistic, because from the point of view of a chosen observer, the idea of "properties of a physical system" makes perfect sense (to the limit imposed by the Kochen-Specker theorem). Indeed, it has been recently argued that relational quantum mechanics would be an instantiation of "structural realism", where "relations", rather than "objects", are seen to be the basic elements of the ontology \cite{Candiotto2017}.

\vspace{1 \baselineskip}

{\bf Quantum events as relational beables}

A priori, it might seem hard to fit the definition of beables as "elements of reality that do not depend on observation" in the framework of relational quantum mechanics. John Bell certainly intended the beables to be observer-independent, but relational quantum mechanics does not claim the complete arbitrariness of reality neither. 

We have seen earlier that the explicit content of the word "beable" was actually given by the peculiar ontology of a chosen interpretation. In the relational interpretation, the basic elements of physical reality are the "relational quantum events" \citep{Rovelli2018}. A quantum event is an interaction between two quantum systems. In this sense, the relational interpretation gives primacy to "relations" over "objects". A measurement is a special kind of interaction where one of the two systems is macroscopic. Other kinds of quantum events happen when two quantum systems entangle through local interaction, and the degrees of freedom of one become correlated with the degrees of freedom of the other. Importantly, these quantum events are themselves "relational" because their mere existence can only be experienced through their interaction with a reference observer.

In fact, if one sticks to a particular observer, a relational beable is nothing but a quantum event. It will not be as absolute as Bell would have expected, because talking about "quantum events" still requires us to first fix a reference observer, but a beable can still be conceived as an "element of reality with respect to the reference observer". Now, given a reference observer $\mathcal{O}$, \textit{the only physically meaningful beables lie in the past cone of $\mathcal{O}$}. Indeed for $\mathcal{O}$, it is a matter of metaphysical faith to attribute an existence to events beyond the scope of its practical experience (future or space-like separated events), but it is a matter of experimental facts to attribute an existence to events in its past cone.

\vspace{1 \baselineskip}

{\bf Relational local causality}

\begin{figure}[h!]
	\includegraphics[width = 1 \columnwidth]{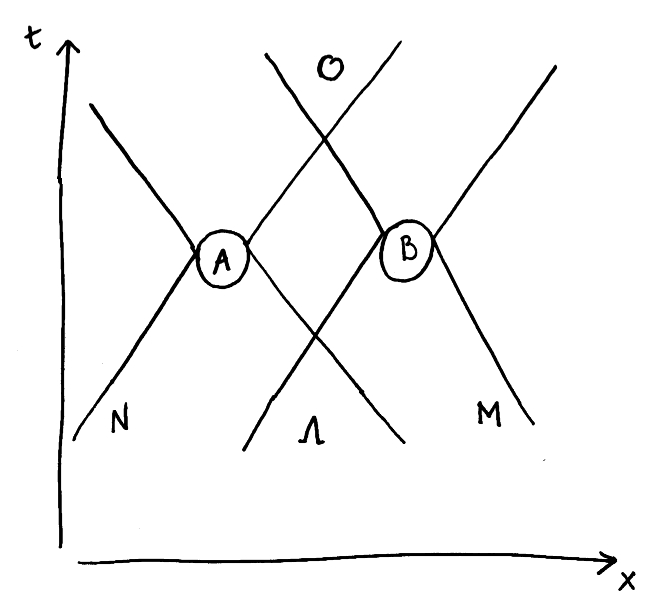}
	\caption{Spacetime diagram showing the localisation of the relational beables $A$,$B$, $\Lambda$, $N$ and $M$, and the observer $\mathcal{O}$ with respect to which they are described.}
	\label{diagram2}
\end{figure}

So, which reference observer $\mathcal{O}$ shall we choose to reformulate the definition of local causality? In order to reasonably talk of the beables $A$ and $B$, the reference observer $\mathcal{O}$ should lie in the common future of $A$ and $B$. (see Figure \ref{diagram2}).

Then the local causality relation should be rewritten
\begin{equation}\label{relational locality}
\left\{ a | \lambda, n, b \right\}_\mathcal{O} = \left\{ a | \lambda, n \right\}_\mathcal{O},
\end{equation}
with an $\mathcal{O}$ index reminding us of the reference observer. A violation of this condition means that, from the perspective of $\mathcal{O}$, $A$ and $B$ are correlated. You can imagine $\mathcal{O}$ as an experimenter looking at two distant Geiger counters, in $A$ and $B$, equipped with a LED which lights up when they detect an $\alpha$ particle. $\mathcal{O}$ would observe that the light coming from $A$ is switched off whenever the light from $B$ is switched on, and vice versa. Now, suppose $\mathcal{O}$ also knows all the relevant beables in region $\Lambda$ and $N$, then he is still unable to predict which of the two Geiger counters will detect the $\alpha$ particle, but he observes, indeed, correlations between $A$ and $B$, because for consistency, there can be at most one $\alpha$ particle emission. This is the only observational manifestation of "non-locality".

\vspace{1 \baselineskip}

{\bf A common cause}

Since $A$ and $B$ are space-like separated, a classical observer $\mathcal{O}$ observing correlations between past events $A$ and $B$ would be lead to the conclusion that there is a common cause to $A$ and $B$. And indeed there is one, namely the radioactive particle in their common past $\Lambda$. 
Here the notion of "common cause" does not entail a deterministic evolution for the decay of the radioactive particle. In the classical case, the notion of causation is strongly related to that of determinism, hence the idea that a complete knowledge of the past would entail a complete knowledge of the future. In quantum mechanics, there is no such a determinism, not only because there is no precise meaning of "complete knowledge of the past" (Heisenberg uncertainty), but also because the measurement process is intrinsically probabilistic. Nevertheless, the intuitive notion of causation does not disappear from the quantum world: in our example it is still meaningful to say that the radioactive element is the "common cause" of the correlations observed later, even if one cannot predict from the past knowledge ($\Lambda$, $N$ and $M$) whether the $\alpha$ particle will be observed in $A$ or in $B$. To be clear, we are here referring to an intuitive notion of "common cause" and not to the formal notion introduced by Reichenbach \cite{Arntzenius2010}. Indeed, his attempt of formalisation is tied too much to a deterministic context, and so, does not suit the example of a radioactive decay as was first pointed out in \cite{VanFraassen1980}.

\vspace{1 \baselineskip}

{\bf Conclusion}

Let us recall that the initial goal of Bell was to formalise the intuitive idea that "causes and effects cannot go faster than light" in the context of the indeterminism of quantum mechanics. This aim was thought to be achieved with the mathematically well-defined notion of "local causality". With this definition indeed, EPR-type experiments have shown that quantum mechanics is fundamentally non-local. What this really means however depends on the  physical content given to beables. Now we have seen that the relational interpretation of quantum mechanics forces us to reconsider the EPR-type experiments from the perspective of a future observer. As a consequence, the failure of "local causality" in the sense of Bell can nevertheless be understood as the existence of a common cause in an indeterministic context. Surely, there is no need to appeal to a mysterious space-like separated influence to understand it. 

A possible conclusion is that Bell's definition of "local causality" does not capture finely enough the intuitive idea of an indeterministic "no superluminal causal influence" as Bell would have liked. Interestingly, Bell himself seems to have been very conscious of the potential inadequacy of his formalisation. Just before asserting his definition of "local causality", Bell writes very honestly:
\begin{quote}
\textit{Now it is precisely in cleaning up intuitive ideas for mathematics that one is likely to throw out the baby with the bathwater. So the next step should be viewed with the utmost suspicion.} \citep{Bell1990}
\end{quote}
With the relational interpretation, quantum mechanics is still "non-local" in the sense of Bell, but it nonetheless remains true that "causes and effects of events are no further away than permitted by the velocity of light". Indeed, if the observer $\mathcal{O}$ sees a light signal from $A$, he will think the radioactive particle in the region of the past cone $\Lambda$ is the cause of the detection of an $\alpha$ particle in $A$. The same thing could be said symmetrically for $B$. Neither causes nor effects travel faster than light whatsoever. There are correlations between $A$ and $B$ because there is a common cause in their common past.

Earlier claims in \cite{Smerlak2007} that relational quantum mechanics was local were maybe misleading: the relational interpretation is not locally causal in the sense of Bell. However it should be clear now that this kind of non-locality cannot be interpreted as a superluminal interaction, and the relational interpretation is indeed local in all the senses listed in section \ref{1}.

Bell did not know about relational quantum mechanics, but he made very clever claims about realism in \cite{Bell1990}. We have already recalled the enlightening examples about the gauge potential and the British sovereignty, and Bell uses them to show that conventions can always travel faster than light. Though it is not clearly stated in these terms, the idea is already almost there that the question "what cannot go faster than light?" might be a relevant criterion to determine what should be considered as physically real and what should not. With this point of view, the failure of non-locality could have already been reinterpreted by Bell as the impossibility to attribute a relevant physical existence to $B$ from the perspective of $A$, and vice versa. This conclusion would have put him on the way to a relational interpretation, with the idea that one should rather consider a future observer $\mathcal{O}$ to talk consistently of $A$ and $B$.

With the relational interpretation, the possible weirdness of non-local experiments boils down to the weirdness of indeterminism. Surely, fundamental randomness is a characteristic feature of quantum physics; the future is not completely predictable, even in principle. Although it contradicts the prejudices of classical physics, randomness has been much easier accepted in the literature than non-locality. A reason may be that the uncertainty of any measurement already constrains classical physics to be indeterministic in practice. The shift is that classical randomness is epistemic (lack of knowledge) while quantum randomness is fundamental (irreducible indeterminism).  In fact, "non-locality" exemplifies the difficulty to understand together "causality" and "indeterminism" in the same conceptual and mathematical framework.

The relational interpretation has not yet answered all the intriguing questions raised by quantum mechanics \cite{Laudisa2017b}. However, a lot of work has already been achieved in the recent years, which accounts for an increasing interest in the approach \cite{Dorato2016, Hohn2017a}. Indeed, we believe this interpretation is a very promising framework to think about the essential features of quantum physics, as we have shown with the example of non-locality in this article.

\section*{Acknowledgements}

The ideas presented in this paper emerged during the Rethinking Workshop 2018 organised with the support of the Basic Research Community for Physics (BRCP). We thank especially Flavio Del Santo, Johannes Kleiner and Robin Lorenz for useful discussions. Daniel Martinez is also thanked for proofreading. We acknowledge the OCEVU Labex (ANR-11-LABX-0060) and the A*MIDEX project (ANR-11-IDEX-0001-02) funded by the "Investissements d'Avenir" French government program managed by the ANR.

\vfill

\bibliographystyle{utcaps}
\bibliography{/home/pmd/Documents/Bibtex/locality}

\end{document}